\documentclass[aps,prc,reprint,amsmath,amssymb,showpacs,superscriptaddress]{revtex4-2}
\usepackage{CJK}
\usepackage{graphicx}
\usepackage{dcolumn}
\usepackage{bm}
\usepackage{color}
\usepackage{orcidlink}
\usepackage{hyperref}

\begin{document}

\begin{CJK*}{UTF8}{}
\title{Zemach radii and nuclear structure effects in hyperfine splitting of lithium isotopes}
\author{Yilong Yang\orcidlink{0000-0002-5065-1309} ({\CJKfamily{gbsn}杨一龙})}
\affiliation{State Key Laboratory of Nuclear Physics and Technology, School of Physics, Peking University, Beijing 100871, China}

\author{Evgeny Epelbaum\orcidlink{0000-0002-7613-0210}}
\affiliation{Ruhr-Universit\"at Bochum, Fakult\"at f\"ur Physik und Astronomie, Institut f\"ur Theoretische Physik II, D-44780 Bochum, Germany}

\author{Chen Ji\orcidlink{0000-0002-4849-480X} ({\CJKfamily{gbsn}计晨})}
\affiliation{Key Laboratory of Quark and Lepton Physics, Institute of Particle Physics, Central China Normal University, Wuhan 430079, China}
\affiliation{Southern Center for Nuclear-Science Theory, Institute of Modern Physics, Chinese Academy of Sciences, Huizhou 516000, China}

\author{Pengwei Zhao\orcidlink{0000-0001-8243-2381} ({\CJKfamily{gbsn}赵鹏巍})}
\email[Contact author: ]{pwzhao@pku.edu.cn}
\affiliation{State Key Laboratory of Nuclear Physics and Technology, School of Physics, Peking University, Beijing 100871, China}

\begin{abstract}
Nuclear structure effects are essential for describing hyperfine splittings from high-precision atomic spectroscopy measurements. 
These effects are often parametrized by the effective or elastic Zemach radii, with their difference being poorly understood.  
We show that the long-standing discrepancy between the effective and elastic Zemach radii in $^6$Li and $^7$Li could be resolved by taking into account nuclear polarizability effects in \emph{ab initio} nuclear structure calculations. 
Our results demonstrate that nuclear polarizability effects, negligible in ${}^7$Li, dominate in ${}^6$Li and explain the observed significant deviation between the effective and elastic Zemach radii. 
Furthermore, we show that the ratios between the nuclear polarizability contributions in different nuclei are universal in the limit of closure and SU(4) symmetry of nuclear forces. 
In particular, the nuclear polarizability contribution in an odd-odd nucleus is enhanced by a factor of $\mu_p/(\mu_p+\mu_n)\simeq 3$, with $\mu_{n,p}$ denoting the nucleon magnetic moments, compared to its odd-$A$ isotopes. 
The same mechanism also explains the Zemach radius deviations observed in ${}^2$H and ${}^3$He. These findings establish nuclear polarizability as the dominant source of isotope-dependent nuclear corrections to hyperfine splitting in light atoms.
\end{abstract}

\maketitle
\end{CJK*}


\textit{Introduction.} High-precision spectroscopy of atoms has fostered productive intersections among atomic, nuclear, and particle physics by enabling stringent tests of bound-state quantum electrodynamics (QED)~\cite{Yan1995Phys.Rev.Lett.74.4791, Yan1998Phys.Rev.Lett.81.774}, probing nuclear structure~\cite{Lu2013Rev.Mod.Phys.85.1383, Pohl2016Science669, Krauth2021Nature, Pachucki2024Rev.Mod.Phys.96.015001}, and searching for signals of physics beyond the standard model~\cite{Chupp2019Rev.Mod.Phys.91.015001}. 
In particular, hyperfine splitting (HFS), an effect dominantly driven by short-range interactions between the nuclear and lepton magnetic moments, serves as a sensitive probe of the nuclear magnetization structure.

The nuclear structure effect on HFS is driven by the two-photon exchange (TPE) between a nucleus and a lepton.
This effect accounts for the difference between the experimentally observed and QED-predicted HFS of the atom, characterized by the effective Zemach radius $r_Z^{\rm eff}$.
Its elastic component is dominated by the Zemach radius $r_Z$ of the nucleus~\cite{Zemach1956PhysRev.104.1771}, which can be extracted from the nuclear electromagnetic densities.
The difference between $r_Z^{\rm eff}$ and $r_Z$ remains poorly understood in theory.

A notable and unexplained discrepancy between $r_Z^{\rm eff}$ and $r_Z$ exists in $^6$Li.
$r_Z^{\rm eff}$, extracted from HFS measurements of the $2S_{1/2}$ states in neutral $^6$Li atoms~\cite{Puchalski2013Phys.Rev.Lett.243001} and the $2^3S_1$ states in $^6$Li$^+$ ions~\cite{Qi2020Phys.Rev.Lett.183002, Sun2023Phys.Rev.Lett.103002}, yields a much smaller value than $r_Z$ extracted from the experimental root-mean-square (rms) charge and magnetic radii of the $^6$Li nucleus, under the assumption of a Gaussian nuclear density distribution~\cite{Yerokhin2008PhysRevA.78.012513}.

This discrepancy is particularly striking given that the two extractions, $r_Z^{\rm eff}$ from HFS and $r_Z$ from nuclear rms radii, yield nearly the same value for the other stable lithium isotope, $^7$Li~\cite{Puchalski2013Phys.Rev.Lett.243001, Qi2020Phys.Rev.Lett.183002}.
Moreover, even within atomic determinations alone, it is puzzling that $r_Z^{\rm eff}$ of $^6$Li is about $40\%$ smaller than that of $^7$Li, reversing the trend in their charge radii.

Two main factors may underlie this discrepancy.  
First, the nuclear-side extraction of $r_Z$ relies on assuming a Gaussian form for the nuclear density distribution~\cite{Yerokhin2008PhysRevA.78.012513}.
Second, inelastic TPE effects, i.e., nuclear polarizability arising from virtual nuclear excitations, contribute to $r_Z^{\rm eff}$ but not to $r_Z$.
They are known to have non-negligible contributions to the Lamb shift~\cite{Ji2013Phys.Rev.Lett.111.143402, Hernandez2014Phys.Lett.B736.344, NevoDinur2016Phys.Lett.B755.380,Ji2018JPG45.093002} and isotope shifts~\cite{Pachucki2011Phys.Rev.Lett.106.193007, Muli2025Phys.Rev.Lett.134.032502} in light atoms. 
For HFS, to date, nuclear polarizability effects have been studied only in light nuclei with mass number $A \leq 3$~\cite{Khriplovich2004J.Exp.Theor.Phys.98.181, Kalinowski2018Phys.Rev.A.98.062513, Ji2024Phys.Rev.Lett.042502, bonilla2025, Friar2005Phys.Lett.B68,Friar2005Phys.Rev.C014002}. 
For lithium isotopes, however, no such calculations exist, largely due to the challenge of accurately modeling six- and seven-body nuclear wave functions within an \emph{ab initio} framework. 

In this Letter, we perform \emph{ab initio} calculations of both $r_Z$ and $r_Z^{\rm eff}$ for $^{6,7}$Li using a deep-learning variational Monte Carlo (VMC) framework with a novel neural-network wave function, based on nuclear interactions derived in chiral effective field theory (EFT) at next-to-next-to-leading order (N$^2$LO).
Our results reveal that, while nuclear polarizability effects are negligible in $^7$Li, they play a crucial role in $^6$Li, accounting for the observed discrepancy between $r_Z^{\rm eff}$ extracted from HFS and $r_Z$ from nuclear experiments.
Furthermore, we show that the relative magnitude of nuclear polarizability contributions across various light nuclei can be understood in terms of the approximate SU(4) symmetry of nuclear forces, providing a unified explanation for similar anomalies observed in other nuclei, such as $^2$H and $^3$He.

\textit{Neural-network VMC.} We compute the (effective) Zemach radii using nuclear ground-state wave functions provided by VMC calculations.
Conventional VMC methods with Jastrow-type wave function \emph{Ans\"atze} provide only approximate solutions for few-nucleon systems, with accuracy deteriorating rapidly for larger $A$ due to the limited expressiveness of the \emph{Ans\"atze}~\cite{Carlson2015Rev.Mod.Phys.1067}.
Motivated by the universal approximation theorem~\cite{Hornik1989NeuralNetworks359}, neural-network wave function \emph{Ans\"atze} have the potential to overcome this bottleneck~\cite{Carleo2017Science602606}.
Their prior applications in nuclear systems have largely been limited to simplified Hamiltonians omitting the essential pion-exchange interactions~\cite{Adams2021Phys.Rev.Lett.022502, Yang2022Phys.Lett.B137587, Lovato2022Phys.Rev.Research043178, Yang2023Phys.Rev.C034320, Fore2023Phys.Rev.Res.033062, Gnech2024Phys.Rev.Lett.142501}.
Only recently have efforts been made to develop neural-network \emph{Ans\"atze} suitable for high-precision nuclear interactions~\cite{Yang2025ChinesePhys.Lett.051201,yang2025arXiv,wen2025arXiv}. 

We devise here a novel neural-network wave function that enables accurate variational solutions of nuclear ground states with high-precision nuclear interactions:
\begin{equation}\label{eq.ansatz}
|\Psi\rangle=\prod_{n=1}^{n_J}\left[1+\sum_{i<j}\left(U^{(n)}_{ij}+\sum_{k\neq i,j}F^{(n)}_{ij;k}\right)\right]|\Phi\rangle_{IM}.
\end{equation}
It comprises a sequence of $n_J$ correlators incorporating two- and three-body correlations and an antisymmetric part $|\Phi\rangle_{IM}$ controlling the quantum number and the long-range behavior.
The correlations take the form
\begin{equation}\label{eq.corr}
    \begin{split}
        U^{(n)}_{ij}&=\sum_{p=2}^6 u^{(n)}_p(r_{ij};\{r_{ik},r_{jk}\}_{k\neq i,j}) O^p_{ij},\\
        F^{(n)}_{ij;k}&=\sum_{q=1}^8 f^{(n)}_q(r_{ij};r_{ik},r_{jk})\tilde{O}_{ij;k}^q,
    \end{split}
\end{equation}
with $O^p_{ij}$ and $\tilde{O}_{ij;k}^q$ consisting of all the linearly independent local operators built from the spin-isospin Pauli matrices of the nucleon pair $(i,j)$ that are allowed by the symmetries~\cite{Supp}.
In the present calculations, $n_J=4$ is taken to reach the desired accuracy.

For $^{6,7}$Li, $|\Phi\rangle_{IM}$ adopts a cluster structure, partitioning into a $^4$He cluster ($\{\alpha\}$) and the remaining $A-4$ valence $p$-shell nucleons ($\{v\}$),
\begin{equation}\label{eq.phi}
    |\Phi\rangle_{IM}=\mathcal{A}\left\{\sum_{LS} w_{LS} F_{LS}(\bm r_{\{\alpha\}},\bm r_{\{v\}})\Phi_{LSIM}(\{\alpha\};\{v\})\right\}.
\end{equation}
It is antisymmetrized over all partitions and summed over several $LS$ couplings with associated weights $w_{LS}$.
$\Phi_{LSIM}(\alpha;v)$ is a single-particle wave function with orbital angular momentum $L$ and intrinsic spin $S$ coupled to total angular momentum $I$ and azimuthal projection $M$~\cite{Carlson2015Rev.Mod.Phys.1067}.
The correlation functions, $u^{(n)}$, $f^{(n)}$, and $F_{LS}$, are parametrized by neural networks~\cite{Supp}.  

The neural-network wave function is optimized by VMC calculations that minimize the energy expectation value using the stochastic reconfiguration method~\cite{Sorella2005Phys.Rev.B241103, Yang2023Phys.Rev.C034320}.
We employ several nuclear Hamiltonians: the phenomenological AV8$'$+UIX$'$ force~\cite{Pudliner1997Phys.Rev.C1720}, which is a simplified version of the AV18+UIX force~\cite{Pudliner1997Phys.Rev.C1720}, and the local N$^2$LO chiral forces, N$^2$LO(1.0) and N$^2$LO(1.2)~\cite{Gezerlis2013Phys.Rev.Lett.032501,Lynn2016Phys.Rev.Lett.062501,Lynn2017Phys.Rev.C054007}.
The optimization procedure and convergence pattern of the ground-state energy with increasing $n_J$ are provided in Ref.~\cite{Supp}.

The enhanced efficiency of our wave function \emph{Ans\"atze} [Eq.~(\ref{eq.ansatz})] stems from two key innovations: the usage of multiple correlators in sequence, and the inclusion of three-body spin-isospin-dependent operators with neural-network-parametrized correlation functions.

\begin{figure}[!htbp]
    \centering
    \includegraphics[width=1\linewidth]{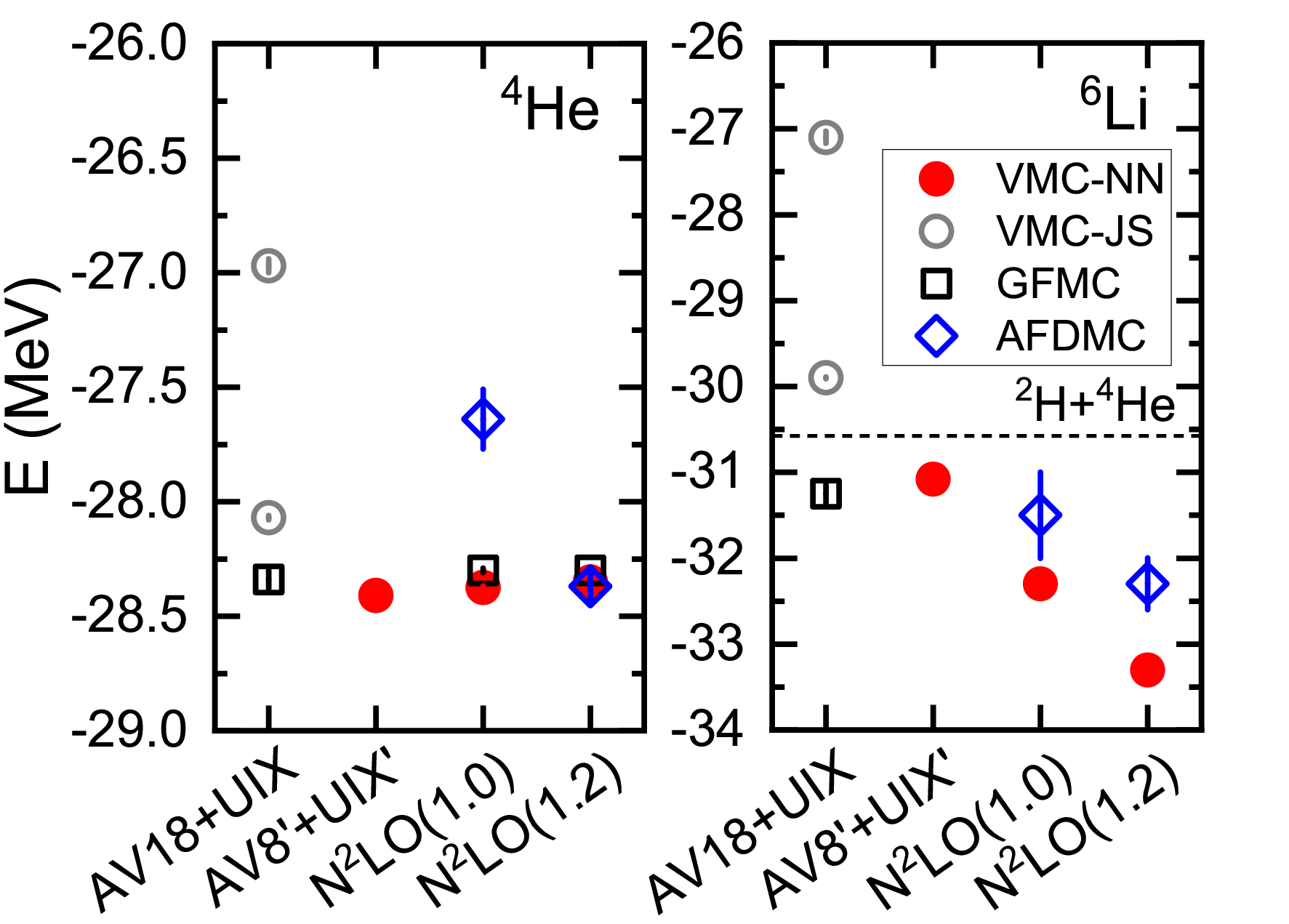}
    \caption{Ground-state energies of $^4$He and $^6$Li computed with the VMC-NN method, compared to the conventional Jastrow-based VMC (VMC-JS)~\cite{Wiringa2000Phys.Rev.C014001, Usmani2012Phys.Rev.C034323}, Green's function Monte Carlo (GFMC)~\cite{Pudliner1997Phys.Rev.C1720,Lynn2016Phys.Rev.Lett.062501}, and auxiliary-field diffusion Monte Carlo (AFDMC)~\cite{Lonardoni2018Phys.Rev.C044318} results.
    The error bars denote Monte Carlo statistical errors.}
    \label{fig1}
\end{figure}
Figure~\ref{fig1} benchmarks the accuracy of the neural-network VMC (VMC-NN) method against previous quantum Monte Carlo results for the $^4$He and $^6$Li ground-state energies.
The conventional Jastrow-based VMC (VMC-JS)~\cite{Wiringa2000Phys.Rev.C014001}, routinely used to provide initial wave functions for Green's function Monte Carlo (GFMC), gives energies of about $-27$ MeV for both $^4$He and $^6$Li; its most sophisticated available extension still leaves $^6$Li above the $^4$He+$^2$H breakup threshold~\cite{Usmani2012Phys.Rev.C034323}.
Relative to the best VMC-JS results, VMC-NN lowers the energies by 0.3 MeV for $^4$He and 1.1 MeV for $^6$Li, reaching an accuracy on par with GFMC, and even outperforms auxiliary-field diffusion Monte Carlo (AFDMC) in some cases.
Crucially, VMC-NN resolves the long-standing issue in conventional VMC, where the $^6$Li ground state tends to unphysically break into $^4$He+$^2$H subclusters.
As a result, the present method yields a meaningful variational minimum with realistic rms radii, which are often not achievable in conventional VMC~\cite{Wiringa2000Phys.Rev.C014001}, and good descriptions of electromagnetic form factors~\cite{Supp}.

\textit{Elastic Zemach radii.} For the study of Zemach radii, we employ the local chiral N$^2$LO(1.0) and  N$^2$LO(1.2) interactions~\cite{Gezerlis2013Phys.Rev.Lett.032501,Lynn2016Phys.Rev.Lett.062501,Lynn2017Phys.Rev.C054007} so that the chiral-truncation uncertainty can be estimated.
With VMC-NN wave functions, the charge ($\rho_E$) and magnetic ($\rho_M$) density distributions are computed to determine the Zemach radii~\cite{Zemach1956PhysRev.104.1771}, 
\begin{equation}\label{eq.rZ}
	 r_{Z}= \int d^3r\, d^3r' \rho_E(\bm r) \rho_M(\bm r')|\bm r'-\bm r|.
\end{equation}
The densities are normalized and are defined as~\cite{Jackson1999}
\begin{equation}
    \rho_E(\bm r)=\frac{1}{Z}\langle 0|\rho(\bm r)|0\rangle,\quad \rho_M(\bm r)=\frac{1}{\mu}\langle 0|\frac{1}{2}[\bm r\times\bm j(\bm r)]_z|0\rangle
\end{equation}
with $|0\rangle$ the nuclear ground state with maximum spin projection ($M=I$) and $\mu$ the nuclear magnetic moment.
We employ the charge density and current operators ($\rho$ and $\bm j$) in the impulse approximation.
The two-body currents are not included because the leading two-body current is found to have minor impact on $r_Z$~\cite{Supp}.

\begin{figure}[!htbp]
    \centering
    \includegraphics[width=0.9\linewidth]{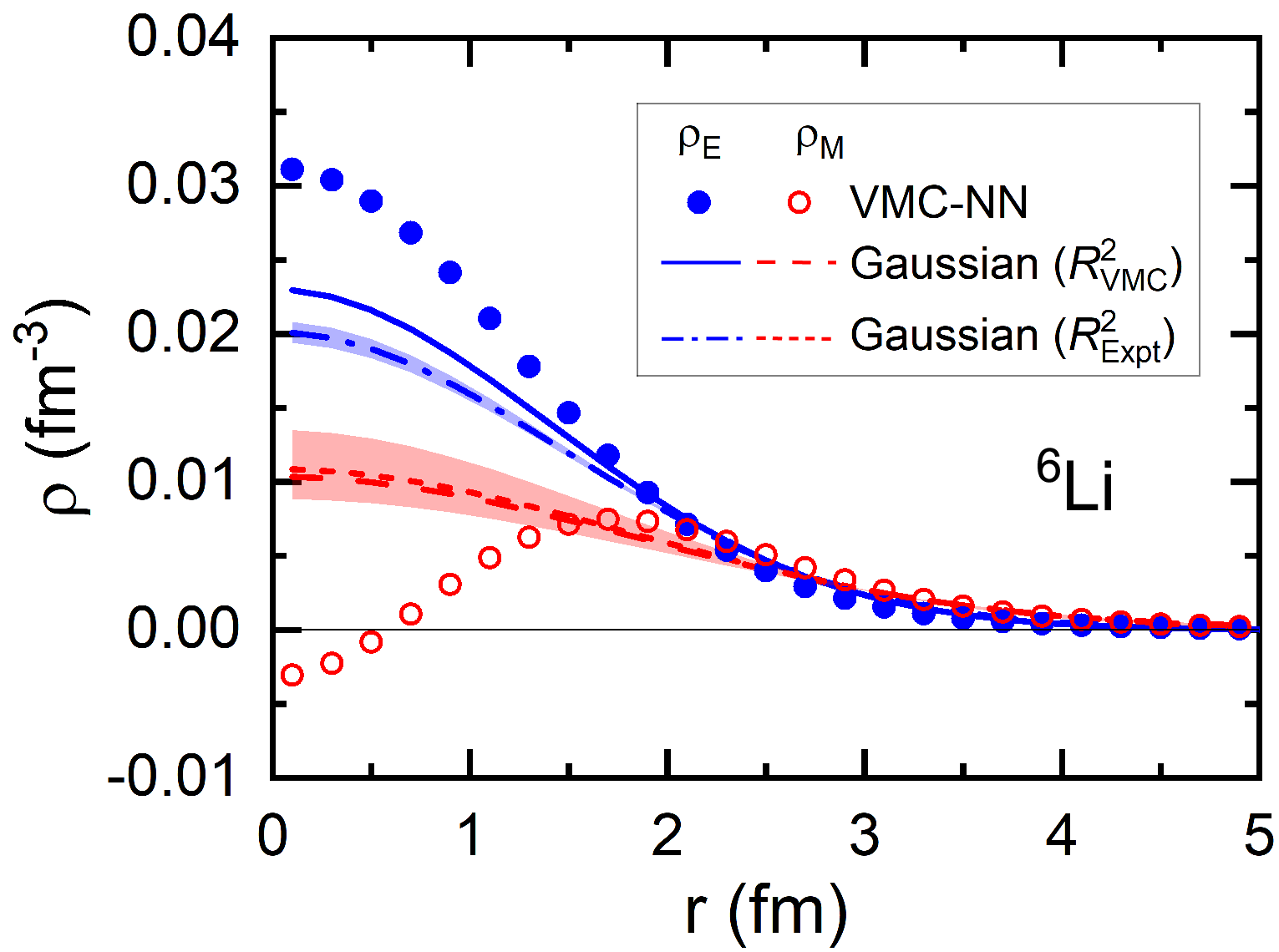}
    \caption{
    The $^6$Li densities $\rho_{E,M}$ obtained with the VMC-NN calculations using the chiral N$^2$LO(1.0) force, compared to the Gaussian densities fixed by the calculated rms radii $(R^2_{\rm VMC})$ and by the experimental rms radii $(R^2_{\rm Expt.})$~\cite{Yerokhin2008PhysRevA.78.012513}.
    }
    \label{fig2}
\end{figure}
Figure~\ref{fig2} depicts the $^6$Li densities $\rho_{E,M}$ obtained with the VMC-NN calculation using the chiral N$^2$LO(1.0) force and compares them with the Gaussian distributions fixed by the rms radii obtained with the same calculation.
Notably, the Gaussian distributions deviate substantially from the realistic VMC-NN densities at short distances ($r \lesssim 2$ fm).
A similar pattern is observed for $^7$Li.

Although the density distributions themselves are not Gaussian in shape, the density differences occur mainly at short distances, where the $4\pi r^2$ metric suppresses their contribution to $r_Z$ [Eq.~\eqref{eq.rZ}].
As listed in Table~\ref{tab1}, the Zemach radii from Gaussian densities ($R_{EM}$) closely match those computed from the full, realistic distributions ($r_Z$), ruling out the detail of nuclear density shapes as the discrepancy source between $r_Z^{\rm eff}$ and $r_Z$ in $^6$Li.
The N$^2$LO(1.2) interaction yields a smaller Zemach radius than the N$^2$LO(1.0) interaction, consistent with the binding energy results (Fig.~\ref{fig1}), but both interactions yield $R_{EM}$ close to $r_Z$.

\begin{table}[!htbp]
    \centering
    \caption{Magnetic moment $\mu$, charge radius $R_E$, magnetic radius $R_M$, and Zemach radius $r_Z$ of $^6$Li.
$R_{EM}$ denotes the Zemach radius estimated with $R_E$ and $R_M$ assuming Gaussian densities~\cite{Yerokhin2008PhysRevA.78.012513}.
For the chiral N$^2$LO(1.0) and  N$^2$LO(1.2) forces, $R_E$ and $R_M$ are computed from the corresponding density distributions, $R_x^2=\int{\rm d}^3 r \ r^2\rho_x(\bm r)$, $x=E,M$.
}
    \label{tab1}
    \begin{tabular}{lccccc}
    \hline\hline
    $^6$Li           & $\mu$    &  $R_E$  & $R_M$    & $R_{EM}$ & $r_Z$   \\
                     & $\mu_N$  &  (fm)   & (fm)     & (fm)     & (fm)    \\\hline
    N$^2$LO(1.0)     & 0.799(1) & 2.43(1) & 3.17(1)  & 3.69(1)  & 3.65(1) \\
    N$^2$LO(1.2)     & 0.788(1) & 2.26(1) & 2.92(2)  & 3.42(2)  & 3.39(2) \\
    Expt.             & 0.822    & 2.54(3) & 3.12(22) & 3.71(26) & -       \\
    \hline\hline
    \end{tabular}
\end{table}
 
\textit{Nuclear polarizability effects.} The polarizability operator takes the form~\cite{Puchalski2013Phys.Rev.Lett.243001, Ji2024Phys.Rev.Lett.042502}
\begin{equation}\label{eq.Hpol}
    \begin{split}
    H_{\rm pol}&={\rm i}(4\pi\alpha)^2\sum_l\delta(\bm r^{(l)})\bm\sigma^{(l)}\cdot\int\frac{{\rm d}^3 q}{(2\pi)^3}\ \\
        &\times\mathop{{\int}\!\!\!\!\!\!\!\!\sum} \limits_{N\neq 0}\frac{h(\omega_N, q)}{4 q^4}\langle0|\{\bm q\times\bm j(-\bm q), \rho(\bm q)\}|0\rangle,
    \end{split}
\end{equation}
where $\bm r^{(l)}$ is the displacement between the $l$th electron and the nucleus, $\bm\sigma^{(l)}$ the spin of the electron, $\bm j$ and $\rho$ the nuclear current and charge-density operators, and $\omega_N$ the excitation energy of the intermediate state $|N\rangle$. 
The kernel is
\begin{equation}
    h(\omega, q)=\frac{1}{m_e^2}\left[\left(2+\frac{\omega}{E_q}\right)\frac{E_q^2+m_e^2+E_q\omega}{(E_q+\omega)^2-m_e^2}-\frac{2q+\omega}{q+\omega}\right]
\end{equation}
with $m_e$ the electron mass and $E_q=(q^2+m_e^2)^{1/2}$.
Nuclear polarizability effects are derived using the Wigner-Eckart theorem, expressing $H_{\rm pol}$ in terms of the Fermi hyperfine energy $E_F$~\cite{Fermi1930ZeitschriftfuerPhysik320},
\begin{equation}
    H_{\rm pol}=-2Z\alpha m_eE_F\ \!\delta r_{\rm pol}.
\end{equation}
By Fourier transforming to coordinate space, 
\begin{equation}\label{eq.rpol1}
    \begin{split}
         \delta r_{\rm pol}&=-\frac{3}{16\mu Z}\int{\rm d}^3r{\rm d}^3r' \mathop{{\int}\!\!\!\!\!\!\!\!\sum} \limits_{N\neq 0}f(\omega_N, |\bm r'-\bm r|)\\
         &\times \langle0|\{\rho(\bm r),((\bm r-\bm r')\times \bm j(\bm r'))_z\}|0\rangle.
    \end{split}
\end{equation}
Here, the expectation value is taken for the nuclear ground state with spin projection $M=I$, and $f(\omega_N, r)$ is a coordinate-space kernel~\cite{Supp}.
Equation~(\ref{eq.rpol1}) sums over all excited states in the TPE process, which is currently feasible only for $^2$H~\cite{Ji2024Phys.Rev.Lett.042502}.
To go beyond $^2$H, certain approximations have to be adopted to make numerical calculations tractable.
The previous study for $A\leq 3$ systems~\cite{Friar2005Phys.Lett.B68, Friar2005Phys.Rev.C014002} utilizes the Low-term formalism~\cite{Low1950Phys.Rev.77.361}, which takes the limit $\omega_N\rightarrow0$ such that $f(\omega_N,r)\rightarrow r$.
However, this formalism overestimates the nuclear polarizability effects in $^2$H and $^3$He~\cite{Ji2024Phys.Rev.Lett.042502, Friar2005Phys.Rev.C014002}.

In this Letter, we propose an alternative formalism without taking the limit $\omega_N\rightarrow0$.
First, using $(H-E_0)/\omega_N\equiv\hat{1}$ and current conservation $[\rho(\bm r),H]=-{\rm i}\nabla\cdot\bm j(\bm r)$, we rewrite Eq.~(\ref{eq.rpol1}) as
\begin{equation}\label{eq.rpol2}
    \begin{split}
    \delta r_{\rm pol}&=\frac{3{\rm i}}{16\mu Z}\int{\rm d}^3r{\rm d}^3r' \mathop{{\int}\!\!\!\!\!\!\!\!\sum} \limits_{N\neq 0}\frac{f(\omega_N, |\bm r'-\bm r|)}{\omega_N}\\
    &\times\langle0|\left[\nabla\cdot\bm j(\bm r), ((\bm r-\bm r')\times \bm j(\bm r'))_z\right]|0\rangle.
    \end{split}
\end{equation}
We use the closure approximation to replace $\omega_N$ with a state-independent average $\bar\omega$, so that the completeness relation $\int\!\!\!\!\!{\scriptstyle\sum}{}_{N\neq 0}|N\rangle \langle N|=\hat{1}-|0\rangle\langle 0|$ reduces $\delta r_{\rm pol}$ to a ground-state observable. 
The closure approximation is justified because $\delta r_{\rm pol}$ is expected to be dominated by low-energy states of $O(10)$ MeV, as shown by the exact calculations for $^2$H~\cite{Ji2024Phys.Rev.Lett.042502}. 
 Although $\bar\omega$ is in principle not a free parameter, its exact determination requires solving excitation spectra.
 Therefore, we instead adjust it to the polarizability contribution extracted from experiments.

We employ the charge density and current operators in the impulse approximation, which are sums over the single-nucleon charge density $\rho^e_i$ and magnetic density $\rho^m_i$~\cite{Kelly2004Phys.Rev.C068202}, leading to~\cite{Supp}
\begin{equation}\label{eq.clos}
    \delta r_{\rm pol}(\bar{\omega})=-\frac{1}{\mu Z}\sum_{i=1}^A\langle0|\mu_i\delta r_i(\bar{\omega})\sigma_{zi}|0\rangle,
\end{equation}
with  
\begin{equation}
    \begin{split}
      \delta r_i(\bar{\omega})&=\frac{1}{8M\bar{\omega}}\int{\rm d}^3r{\rm d}^3r'\rho^e_i(|\bm r+\bm r'|)\rho^m_i(r')\\
      &\times\nabla^2[3f(\bar{\omega},r)+r\partial_rf(\bar{\omega},r)],
    \end{split}
\end{equation}
$\mu_i$ ($\sigma_{zi}$) the magnetic moment (the Pauli spin operator) of the $i$th nucleon, and $M$ the nucleon mass.

Following Eq.~(\ref{eq.clos}), the relative magnitudes of nuclear polarizability contributions $\delta r_{\rm pol}$ can be qualitatively estimated by invoking the (approximate) SU(4) symmetry~\cite{Wigner1937Phys.Rev.106}.
In the SU(4) limit, the spins of even-numbered protons and neutrons saturate and, thus, only unpaired nucleons dominate the magnetic moment.
Since the neutron has zero total charge and a small charge distribution, $\delta r_n\simeq 0$.
Therefore, $\delta r_{\rm pol}$ is approximately estimated according to proton ($Z$) and neutron ($N$) numbers' parities:
\begin{equation}\label{eq.su4}
    \delta r_{\rm pol}(\bar{\omega})\simeq
    \left\{
    \begin{split}
        &-\frac{1}{Z}\frac{\mu_p}{\mu_p+\mu_n}\delta r_p(\bar{\omega}) &\text{odd } Z, \text{odd } N,\\
        &-\frac{1}{Z}\delta r_p(\bar{\omega}) &\text{odd } Z, \text{even } N,\\
        &0 &\text{even } Z, \text{odd } N,\\
    \end{split}
    \right.
\end{equation}
with $\mu_p=2.793\mu_N$ and $\mu_n=-1.913\mu_N$. 
In even-$Z$ nuclei, $\delta r_{\rm pol}$ is weak due to cancellation between spin-up and spin-down protons. 
For odd-$Z$ nuclei, $\delta r_{\rm pol}$ scales with $Z^{-1}$ and the factor $\mu_p/(\mu_p+\mu_n)\simeq3$ enhances $\delta r_{\rm pol}$ in odd-$Z$ odd-$N$ nuclei over odd-$Z$ even-$N$ ones.

\begin{figure}[!htbp]
    \centering
    \includegraphics[width=0.95\linewidth]{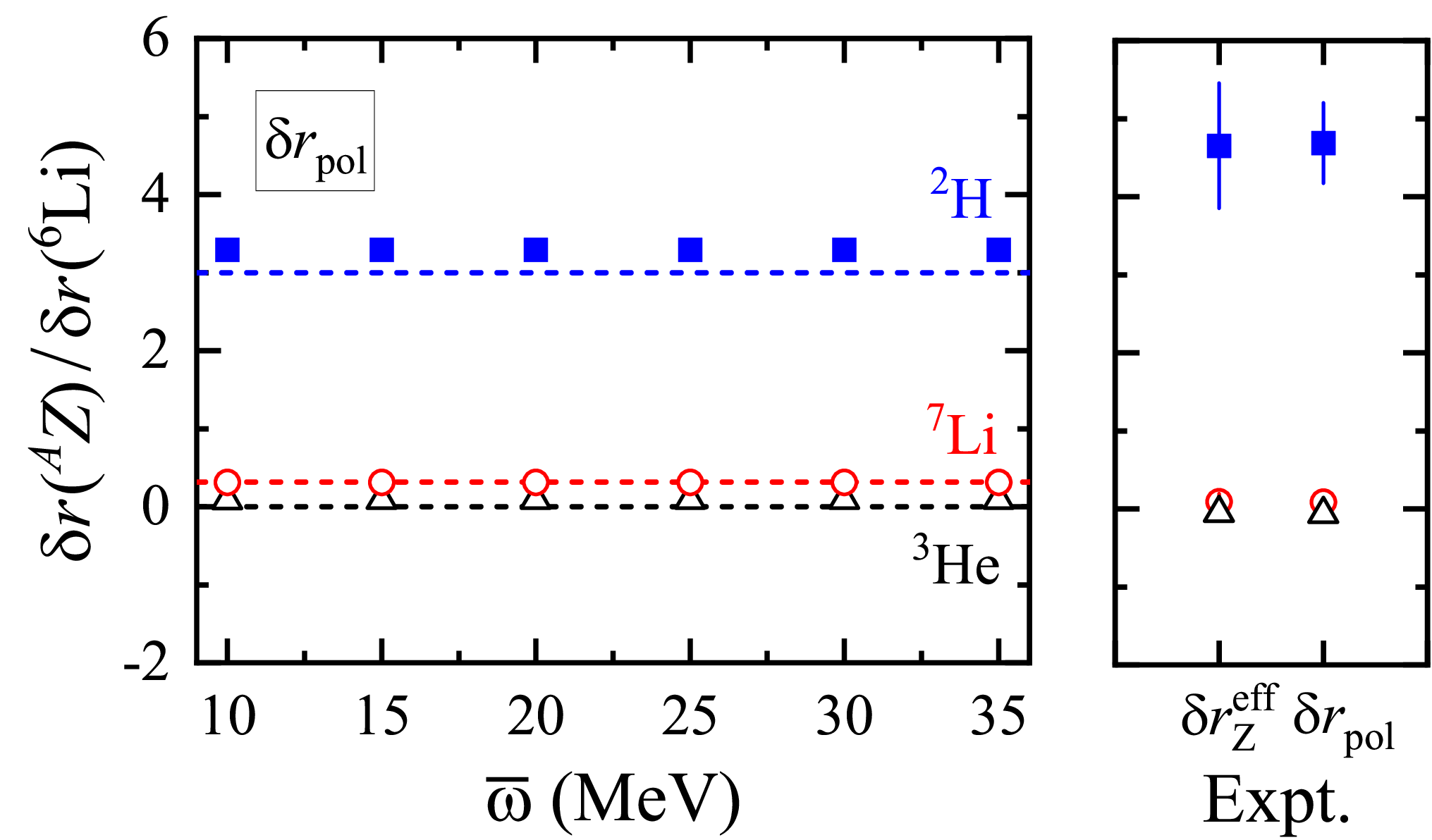}
    \caption{Left panel: Ratios of $\delta r_{\rm pol}$ between various nuclei and $^6$Li, $\delta r_{\rm pol}(^AZ)/\delta r_{\rm pol}({\rm ^6Li})$, as functions of the closure energy $\bar{\omega}$. Dots represent the VMC-NN results with the N$^2$LO(1.0) force and dashed lines denote the SU(4)-limit values [Eq. (\ref{eq.su4})].
    Right panel: Ratios of $\delta r_Z^{\rm eff}=r_Z^{\rm eff}-r_Z$ and $\delta r_{\rm pol}$ between various nuclei and $^6$Li extracted from experiments; see text for details.}
    \label{fig3}
\end{figure}

These trends are confirmed in Fig.~\ref{fig3}, where ratios of nuclear polarizability effects between a nucleus $^A Z$ and $^6$Li, i.e., $\delta r_{\rm pol}(^AZ)/\delta r_{\rm pol}({\rm ^6Li})$, are obtained with the VMC-NN calculations using the chiral N$^2$LO(1.0) force, assuming a uniform closure energy $\bar{\omega}$ across the considered nuclei.
The calculated ratios vary weakly with $\bar{\omega}$ over tens of MeV and align with the SU(4)-limit estimates [Eq.~(\ref{eq.su4})]. 
Notably, the SU(4) estimates of $\delta r_{\rm pol}$ also qualitatively capture the observed discrepancies between $r_Z^{\rm eff}$ extracted from HFS measurements~\cite{Sun2023Phys.Rev.Lett.103002,Puchalski2013Phys.Rev.Lett.243001,Pachucki2023Phys.Rev.A052802,Krauth2016AnnalsofPhysics168,Patkos2023Phys.Rev.A052802} and $r_Z$ determined from experimental nuclear rms radii $R_{E,M}$~\cite{Yerokhin2008PhysRevA.78.012513} or form factors~\cite{Friar2004Phys.Lett.B285,Sick2014Phys.Rev.C064002}. 
In particular, the differences $r_Z^{\rm eff} - r_Z$ are smaller in odd-even nuclei ($^7$Li and $^3$He) than in odd-odd nuclei ($^6$Li and $^2$H).
The difference in $^2$H is larger than that in $^6$Li.
These agreements provide strong evidence that the discrepancies between $r_Z^{\rm eff}$ and $r_Z$ originate from nuclear polarizability. 

Note that the closure approximation does not take into account the presence of excited states and resonances in the $\omega$ region of interest, which may affect the strength of certain states and thereby modify $\delta r_{\rm pol}$. 
The weak dependence on $\overline{\omega}$ in Fig.~\ref{fig3} shows the stability of the present calculations, rather than the proof of validity.

\begin{figure}[!htbp]
    \centering
    \includegraphics[width=0.95\linewidth]{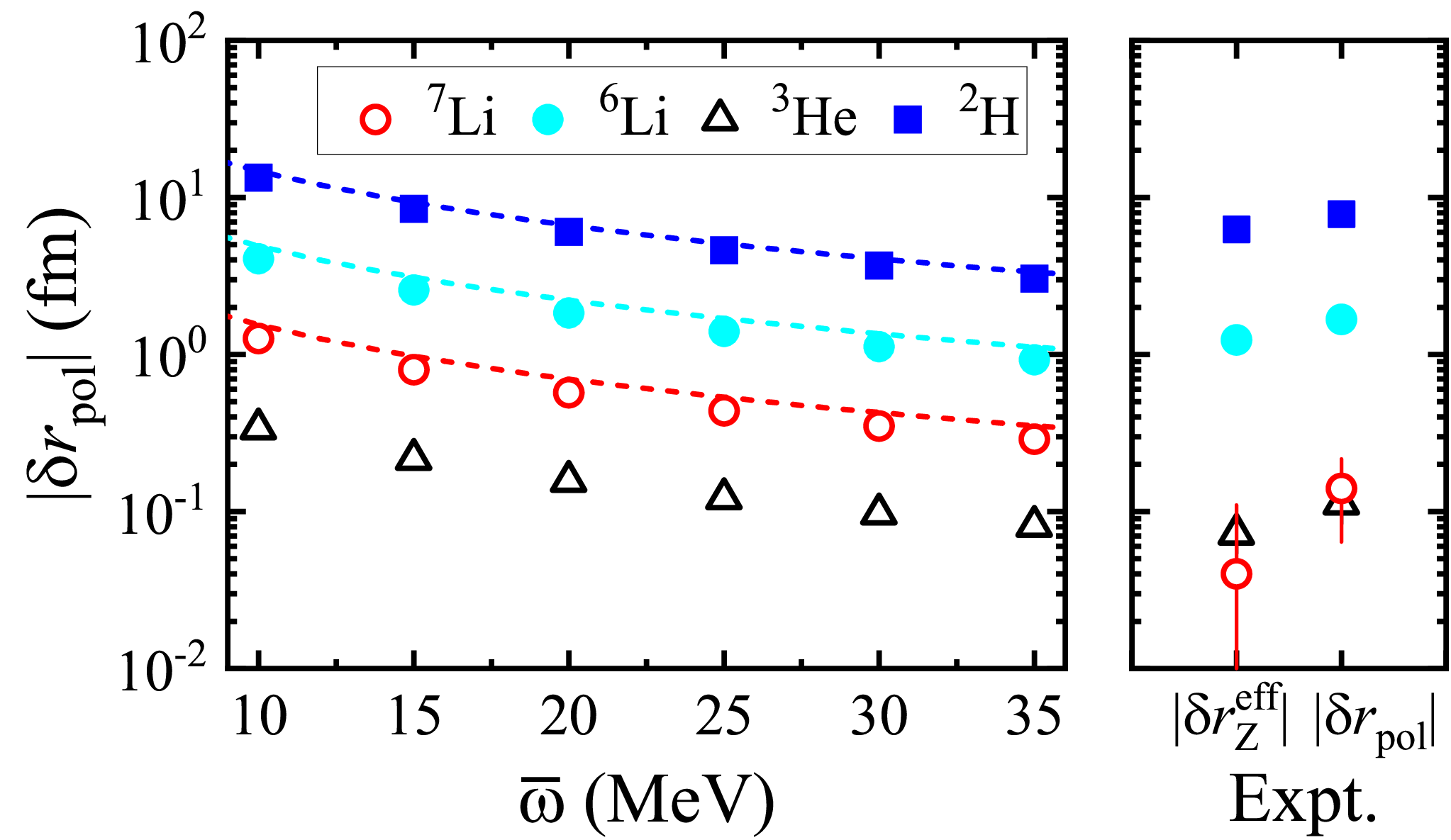}
    \caption{Same as Fig.~\ref{fig3}, but for the magnitudes of $\delta r_{\rm pol}$.}
    \label{fig4}
\end{figure}

In Fig.~\ref{fig4}, the absolute magnitudes $|\delta r_{\rm pol}|$ are shown for various nuclei. 
Although $|\delta r_{\rm pol}|$ for each nucleus varies with $\overline{\omega}$, the relative order of $|\delta r_{\rm pol}|$ in different nuclei is preserved over a wide range of $\overline{\omega}$.

\begin{table}[!htbp]
    \centering
    \caption{The corrections $\delta r_{\rm rec}$ and $\delta r_{\rm 1N}$  obtained from the present VMC-NN calculations with the chiral N$^2$LO(1.0) force.
   Parentheses denote statistical errors.}
    \label{tab.2}
    \begin{tabular}{lcccc}
    \hline\hline
                      & $^2$H      & $^3$He     & $^6$Li     & $^7$Li     \\\hline
    $\delta r_{\rm rec}$ (fm) & $-0.37(1)$ & $-0.25(1)$ & $-0.16(3)$ & $-0.15(3)$ \\
    $\delta r_{\rm 1N}$ (fm)  & $1.94(9)$  & $0.21(2)$  & $0.59(1)$  & $0.25(0)$  \\\hline\hline
\end{tabular}
\end{table}

Besides the Zemach radius and nuclear polarizability, the effective Zemach radius also includes the recoil and single-nucleon TPE contributions~\cite{Ji2024Phys.Rev.Lett.042502},
\begin{equation}
    r_Z^{\rm eff}=r_{Z}+\delta r_{\rm rec}+\delta r_{\rm 1N}+\delta r_{\rm pol}.
\end{equation}
We performed VMC-NN calculations of $\delta r_{\rm rec}$ and $\delta r_{\rm 1N}$~\cite{Supp} and the resulting values are listed in Table~\ref{tab.2}.
Subtracting them from the experimental $r_Z^{\rm eff}-r_Z$ yields the ``experimental'' $\delta r_{\rm pol}$ shown in the right panels of Figs.~\ref{fig3} and \ref{fig4}. 
The small deviations of the SU(4) estimates obtained with a common $\bar{\omega}$ are resolved below using isotope-dependent $\bar{\omega}$.

\begin{figure}[!htbp]
    \centering
    \includegraphics[width=0.95\linewidth]{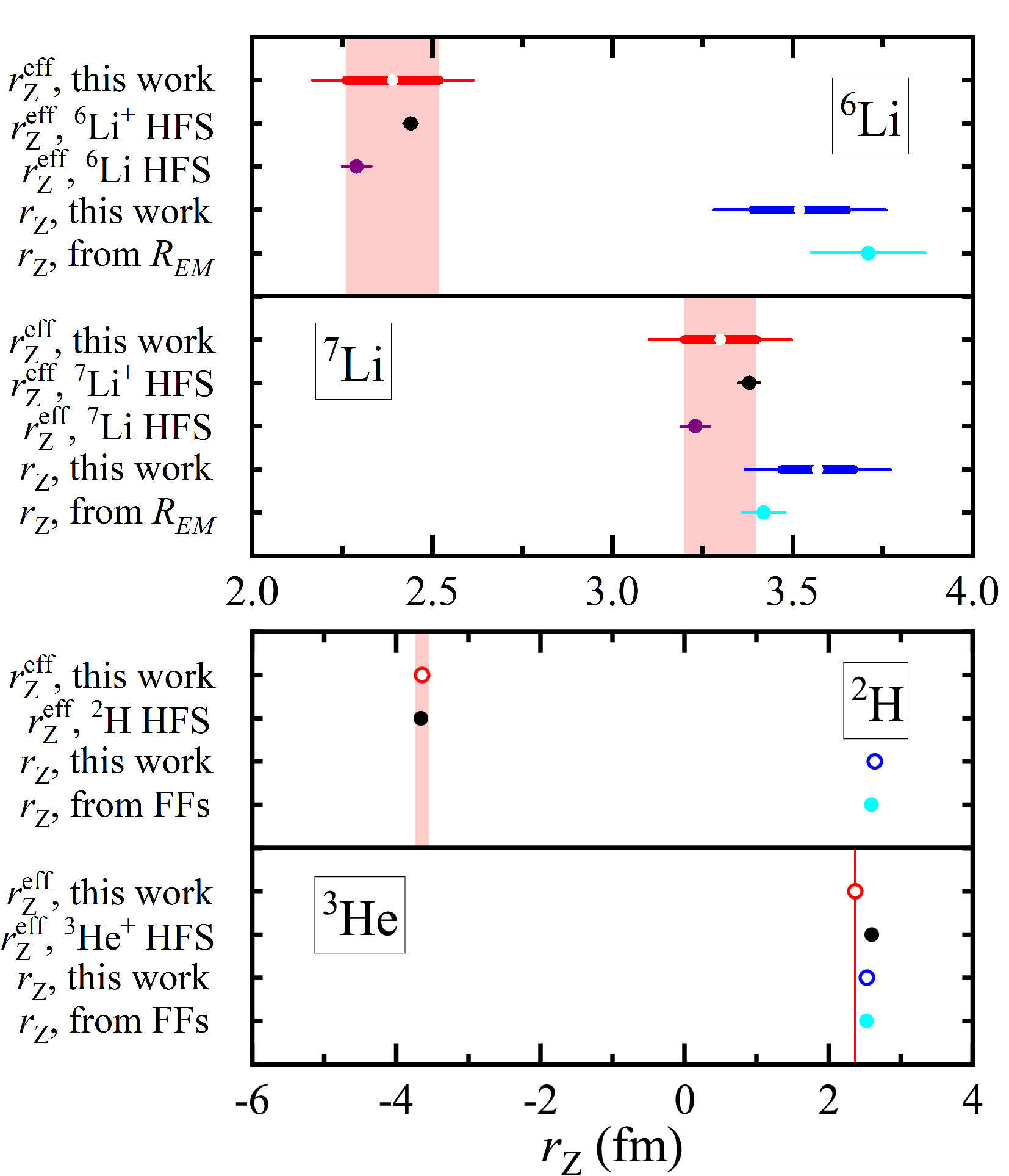}
    \caption{The $r_Z^{\rm eff}$ and $r_Z$ values obtained from the VMC-NN calculations with the chiral N$^2$LO force, compared to those extracted from atomic HFS measurements~\cite{Puchalski2013Phys.Rev.Lett.243001,Sun2023Phys.Rev.Lett.103002,Pachucki2023Phys.Rev.A052802,Krauth2016AnnalsofPhysics168, Patkos2023Phys.Rev.A052802} and nuclear rms radii $R_{EM}$ or form factors~\cite{Yerokhin2008PhysRevA.78.012513,Friar2004Phys.Lett.B285,Sick2014Phys.Rev.C064002}.
    Thick error bars combine Monte Carlo and cutoff-variation uncertainties; thin error bars denote estimated chiral-truncation errors~\cite{Supp}. 
    Uncertainties from the closure approximation are not included.
    }
    \label{fig5}
\end{figure}
In Fig.~\ref{fig5}, the full VMC-NN calculations of $r_Z^{\rm eff}$ for $^{6,7}$Li are compared with the values extracted from atomic HFS measurements~\cite{Puchalski2013Phys.Rev.Lett.243001, Pachucki2023Phys.Rev.A052802, Li2020Phys.Rev.Lett.063002, Qi2020Phys.Rev.Lett.183002, Sun2023Phys.Rev.Lett.103002}.  
The $r_Z^{\rm eff}$ values from HFS are well reproduced by adjusting the closure energies, $\bar{\omega}=23$ MeV for $^6$Li and $\bar{\omega}=29$ MeV for $^7$Li; because the closure energies are adjusted to data, the associated uncertainty from the closure approximation is not estimated.
The calculated $r_Z$ values agree well with experiment, and $\delta r_{\rm pol}$ has the correct sign and order of magnitude to account for $r_Z^{\rm eff}-r_Z$. 
Also, these results remain robust against chiral-truncation uncertainties and two-body current effects~\cite{Supp}.
Therefore, the present work motivates future efforts to explicitly account for nuclear excitations, though technically challenging, to accurately calculate nuclear polarizability contributions in $^{6,7}$Li.

The present framework can also be applied to $^2$H and $^3$He, also shown in Fig.~\ref{fig5}.
The $r_Z^{\rm eff}$ values from HFS are described with $\bar{\omega}=16$ MeV for $^2$H and $\bar{\omega}=25$ MeV for $^3$He.

\textit{Summary.} We report the first \emph{ab initio} calculation of the Zemach radii of $^{6,7}$Li and demonstrate that nuclear polarizability effects dominate the difference between the effective Zemach radii $r_Z^{\rm eff}$ extracted from hyperfine splittings and the elastic Zemach radii $r_Z$ derived from nuclear charge and magnetic radii.
We develop a novel neural-network wave function for VMC calculations that enables accurate variational solutions for $A\leq 7$ ground states with chiral nuclear interactions.
Then, we derive a compact expression for nuclear polarizability effects using the closure approximation and calculate both $r_Z$ and $r_Z^{\rm eff}$ within a unified framework.  
The results show that $\delta r_{\rm pol}$ is significantly larger in $^6$Li than in $^7$Li, explaining the observed difference in $r_Z^{\rm eff}-r_Z$ between $^6$Li and $^7$Li. 

Invoking the nuclear SU(4) symmetry, we reveal the systematic enhancement of nuclear polarizability effects in odd-odd nuclei over odd-$A$ ones, and the enhancement decreases with larger $Z$.
The SU(4) estimate, together with the neural-network VMC calculations, verifies the observed $r_Z^{\rm eff} - r_Z$ variation trend in $^2$H, $^3$He, $^6$Li, and $^7$Li.
These findings advance a unified understanding of nuclear-structure effects in the hyperfine structure of light atoms, and the predicted systematics of nuclear polarizability [Eq.~(\ref{eq.su4})] provide testable benchmarks for future studies of other light nuclei, such as $^{7,9}$Be~\cite{Qi2023Phys.Rev.AL010802,Dickopf2024Nature757}.

\textit{Acknowledgments.} We thank F. Guo, Y. Wang, S. Bacca, and S. Pastore for helpful discussions.
Y.L.Y. thanks Ulf-G. Mei\ss{}ner for the valuable discussions during his visit to Bonn University.
This work has been supported in part by the National Natural Science Foundation of China (Grants No. 12141501, No. 123B2080, No. 12435006, No. 12475117, No. 12175083, No. 12335002, and No. 11805078), by the National Key Laboratory of Neutron Science and Technology Grant No. NST202401016, by the National Key R\&D Program of China (Contracts No. 2024YFA1612600 and No. 2024YFE0109803), by the High-performance Computing Platform of Peking University, by the European Research Council (ERC AdG NuclearTheory, Grant No. 885150),
by the MKW NRW under the funding code NW21-024-A, by JST ERATO (Grant No. JPMJER2304) and by JSPS KAKENHI (Grant No. JP20H05636).

\textit{Data availability.} The data that support the findings of this article are not publicly available. The data are available from the authors upon reasonable request.


%

\end{document}